\documentclass[sigconf]{acmart}

\usepackage{amsmath}

\usepackage{amssymb}

\usepackage{amsfonts}
\usepackage{algorithmic}
\usepackage{graphicx}
\usepackage{textcomp}
\usepackage{xcolor}
\usepackage{caption}
\usepackage[normalem]{ulem}

\AtBeginDocument{%
  \providecommand\BibTeX{{%
    \normalfont B\kern-0.5em{\scshape i\kern-0.25em b}\kern-0.8em\TeX}}}

\setcopyright{none}
\acmYear{}

\acmConference[HotInfra\textquotesingle23]{First Workshop on Hot Topics in System Infrastructure}{June 18, 2023}{Orlando, FL, USA.}
\acmBooktitle{First Workshop on Hot Topics in System Infrastructure (HotInfra\textquotesingle23), June 18, 2023, Orlando, FL, USA} \acmPrice{}
\acmISBN{} \acmDOI{}

\newcommand{\etal}[1]{#1~\textit{et~al.}}

\begin{document}
\title{Sidecars on the Central Lane:\texorpdfstring{\\}{} Impact of Network Proxies on Microservices}

\author{Prateek Sahu}
\email{prateeks@utexas.edu}
\orcid{0009-0000-5569-5856}
\affiliation{%
  \institution{University of Texas at Austin}
  \streetaddress{}
  \city{}
  \state{}
  \country{}
  \postcode{}
}

\author{Lucy Zheng}
\email{lucy.zheng@utexas.edu}
\affiliation{%
  \institution{University of Texas at Austin}
  \streetaddress{}
  \city{}
  \state{}
  \country{}
  \postcode{}
}
\author{Marco Bueso}
\email{mbueso@utexas.edu}
\affiliation{%
  \institution{University of Texas at Austin}
  \streetaddress{}
  \city{}
  \state{}
  \country{}
  \postcode{}
}
\author{Shijia Wei}
\email{shijiawei@utexas.edu}
\orcid{0000-0002-4513-5334}
\affiliation{%
  \institution{University of Texas at Austin}
  \streetaddress{}
  \city{}
  \state{}
  \country{}
  \postcode{}
}
\author{Neeraja J. Yadwadkar}
\email{neeraja@austin.utexas.edu}
\orcid{1234-5678-9012}
\affiliation{%
  \institution{University of Texas at Austin, VMware Research}
  \streetaddress{}
  \city{}
  \state{}
  \country{}
  \postcode{}
}
\author{Mohit Tiwari}
\email{tiwari@austin.utexas.edu}
\orcid{0000-0002-0384-3308}
\affiliation{%
  \institution{University of Texas at Austin}
  \streetaddress{}
  \city{}
  \state{}
  \country{}
  \postcode{}
}

\renewcommand{\shorttitle}{Sidecars on the Central Lane}

\begin{abstract}
Cloud applications are moving away from monolithic model towards 
loosely-coupled microservices designs. Service
meshes are widely used for implementing microservices applications
mainly because they provide a modular architecture for
modern applications by separating operational features from
application business logic. \emph{Sidecar proxies} in service meshes
enable this modularity by applying security, networking, and
monitoring policies on the traffic to and from services. To
implement these policies, sidecars often execute complex chains of
logic that vary across associated applications and end up unevenly
impacting the performance of the overall application. Lack of
understanding of how the sidecars impact the performance of
microservice-based applications stands in the way of building
performant and resource-efficient applications. To this end, we
bring sidecar proxies in focus and argue that we need to deeply
study their impact on the system performance and resource
utilization. We identify and describe challenges in characterizing
sidecars, namely the need for microarchitectural metrics and
comprehensive methodologies, and discuss research directions
where such characterization will help in building efficient service
mesh infrastructure for microservice applications.
\end{abstract}




\maketitle

\section{Introduction}


Cloud applications are widely adopting loosely coupled microservice based solutions~\cite{joyner:arxiv:2020:ripple,zhang:mlarchsys:2020:sinan,gan:mlarchsys:2020:sage,shahrad2019architectural,gan:cal:2018:microservices,gan2019open}.
With this change,
operators of large cloud applications face increasing challenges in
traffic management as these applications scale to  
hundreds, or even thousands of services~\cite{gan2019open,Netflixperf,LyftRuns,zhang2022crisp}.
The complex inter-service communication patterns of these microservices make it challenging to restrict, route, and monitor the generated traffic.
Service meshes are widely adopted~\cite{CNCFSurv,istio-cs,ciliumu-cs,linkerd-cs}
as they provide robust frameworks for
implementing and deploying tools that aid in 
simplifying the traffic management.
Service meshes leverage \emph{sidecar proxies} 
that can be configured to execute inter-service communication policies.
These policies enable security, networking, and observability features such as
endpoint access control, request rate limiting, and logging.

%
However, sidecars end up increasing request latency and result in performance penalties
for large applications since they are designed to
co-locate with each application container. 
Prior research~\cite{zhu2022dissecting} reports 30-185\% increase in latency and 41-92\% overhead in CPU usage for different benchmark applications.
Other studies by leading service mesh vendors~\cite{istioperf,linkerd_benchmark,Cilium_benchmark} note 2-6$\times$  latency overheads of the sidecar proxies. These studies also note that the CPU usage of sidecars ranges from negligible overhead to 0.35 vCPU for every 1000 requests based on the choice of sidecars and request rates.

Thus, it is crucial to understand the performance of sidecar proxies to be able to reason about the performance of microservice applications. However, characterizing the performance implications of sidecars with complex inter-service traffic management policies is challenging, since (a) no defined metrics exist, and (b) no established methodology exists to guide application operators and hardware architects to 
systematically optimize the performance of microservice
applications.
Existing efforts~\cite{sriraman2018mu,gan2019open,parties,zhu2022dissecting,zhang:mlarchsys:2020:sinan} are based on traditional 
metrics like latency and CPU or memory utilization to study microservice performance. 
However, these metrics provide no insights into how sidecar proxies interact with the underlying hardware.
To the best of our knowledge, no prior profiling work has investigated the microarchitectural impacts of sidecar proxies in service mesh infrastructures.
Additionally, we observe significant changes in application latencies depending on
the types and complexity of policies used.
However, studies from service mesh practitioners \cite{istioperf,linkerd_benchmark,Cilium_benchmark} neglect the impact of the diverse and complex set of network policies that are supported~\cite{networkfilters_envoy, HTTPfilters_envoy,NetworkPolicy_cilium}. 

Our work brings sidecars into focus and argues for measuring their impact on application performance. 
In doing so, we highlight the lack of any microarchitectural metrics and omission of diverse network policies as two key challenges in understanding the performance overheads of service-mesh sidecars.
We discuss how characterizing sidecar proxies---with
microarchitectural metrics and a methodology covering
diverse policies with varying complexity---enables
application operators to navigate the performance trade-offs
of service mesh infrastructures;
and opens avenues in software and hardware research 
for microservices infrastructure.

\section{Background}
To restrict, route, and monitor the inter-service
traffic in microservice applications,
service meshes deploy one sidecar proxy
alongside each application process
to mediate the corresponding traffic.
Specifically, sidecars apply policies to the
network traffic,
without modifying the main business logic or incurring any service downtime.
Sidecars use a variety of listeners that implement a set of filters commonly known as filter-chains. 
Default filter-chains usually pass through packets with minimal logging and packet modifications.
However, complex policies may require heavy computations, pattern matching, and request modification.
For example,
mTLS~\cite{mTLSRFC} enables mutually authenticating services by encrypting
the inter-service traffic;
application-layer role-based access control (RBAC)~\cite{envoyRBAC} enables fine-grained access control by
restricting API access from certain services;
request tagging~\cite{IPTagging} enables detailed telemetry collection by augmenting
request headers along the service invocation chain.
Furthermore, sidecar vendors provide operators with not only a long list of common
filters~\cite{networkfilters_envoy, HTTPfilters_envoy,NetworkPolicy_cilium} 
but also the option for writing custom filters~\cite{Wasm—env41:online,Lua—envo8:online}.
This configurability and programmability
make the performance characteristics of a sidecar 
vary significantly across different deployment settings.

\section{Challenges}

With the aforementioned complex policies,
we discuss two major challenges in quantitatively navigating the performance space of employing sidecar proxies with diverse policies. 
Our experiments study the performance characteristic when 
an Envoy proxy~\cite{Envoy} is allocated with different
numbers of vCPUs,
and when it is configured with different traffic policies.
We explore two commonly used policies
(a) RBAC (Role Based Access Control) and (b) IP Tag.
The RBAC policy filters out incoming requests based on their source IP address, whereas the IP Tag policy incorporates specific header values to the HTTP request determined by source IP addresses.

\noindent\textbf{Challenge 1:} Lack of adequate metrics fails to highlight bottlenecks in sidecar proxies accurately.

\begin{figure}[b]
    \centering
    \includegraphics[width=0.95\columnwidth]{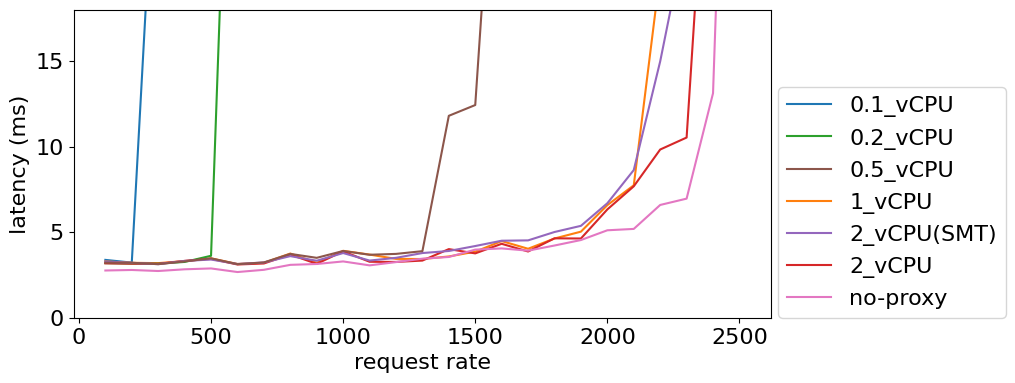}
    \caption{P90 latency for Envoy with increasing 
    vCPU.
    }
    \label{fig:envoy_cpu_latency}
\end{figure}

Existing performance studies of service
meshes~\cite{istioperf,linkerd_benchmark,Cilium_benchmark} focus on 
user-sensitive metrics such as CPU utilization, latency, and throughput.
However, Figure~\ref{fig:envoy_cpu_latency} shows without microarchitectural metrics 
insights like pipeline congestion using top-down analysis,
operators cannot reason about performance improvement with
allocation of additional OS resources.
Figure~\ref{fig:envoy_cpu_latency} plots the latency and throughput
behavior when an Envoy~\cite{Envoy} proxy is allocated various
vCPU time.
We observe that allocation of two virtual cores (threads), mapped to the same physical core provides 
no performance increase over a single core allocated. We however do see throughput improve when we map it to separate cores.
This behavior cannot be explained with system-level metrics and requires detailed
insights into pipeline occupancy and logical unit contention to reason about the observed performance. 

Similarly,
latency and dynamic instruction count of
the IP Tag policy in 
Table~\ref{tab:my_label} show another example of the need of
microarchitectural metrics
for attributing the hardware performance bottleneck.
We notice that increasing the number of tags in IP Tagging 
from 1 to 5, and 10
results in a linear increase in dynamic instructions.
However, over the 1-tag baseline,
cycle counts only increase by 1.8\% for 5 tags yet
by 7\% for 10 tags. 
A deeper investigation reveals that this non-linear overhead is
due to L2-cache misses incurred when processing 10 tags.
IP Tagging with 10 tags shows a nearly 10\% overhead in L2 misses
while processing 5 tags incurs 2.1\%.

\noindent\textbf{Challenge 2:} Neglecting network policies during profiling misleads performance trend prediction for sidecar proxies.

Most of service-mesh performance studies~\cite{smp-meshery,servicemesh-medium} follow practices in
microservices benchmarking~\cite{sriraman2018mu,gan2019open,parties},
which focuses on the impact of request sizes and request rates
on performance metrics over a representative application set.
Although they remain relevant aspects 
to study the performance impact of sidecars, the methodology needs to include diverse policies with varying complexity to provide a comprehensive analysis.
Prior work by Zhu {\em et~al.}~\cite{zhu2022dissecting} 
characterizes the overhead for some filters but omits policy complexities
and their interactions with the microarchitecture.

Table~\ref{tab:my_label} shows that different policies have distinct performance impact on 
application latency, and increasing the policy complexity affects performance differently.
In Table~\ref{tab:my_label}, we note that application latency
for both policies are similar
but the instruction footprint is significantly higher in 
IP tagging. 
Differences in execution profiles and 
the lack of a representative set of filters is a major challenge to predict performance costs. 

\begin{table}
    \centering
    \begin{tabular}{|c|c|c|c|c|} \hline
       Policy & p90 Latency (ms) & p90 Cycles  & Instructions \\ \hline \hline
       IP Tag (1) & 1.034 & 1.035 & 1.052\\ \hline
       IP Tag (5) & 1.039 & 1.054 & 1.088\\ \hline
       IP Tag (10) & 1.048 & 1.108 & 1.147\\ \hline
       RBAC (100) & 1.029 & 1.123 & 1.014 \\ \hline
       RBAC (10k) & 1.044 & 1.137 & 1.014 \\ \hline
    \end{tabular}
    \caption{Latency, cycle, and instruction overhead for two policies (normalized to baseline with no policies) - HTTP request tagging with 1, 5, 10 tags , Role-Based Access Control (RBAC) with 100 and 10k rules respectively.
    }
    \label{tab:my_label}
    \vspace{-33pt}
\end{table}

\section{Research Directions}

By design, sidecars are central places to implement and
consolidate
common operational tasks, also known as the `datacenter tax'.
\etal{Kanev}~\cite{googledctax} suggested that this datacenter tax,
such as protocol management, remote procedure calls, 
and data movement contributes over 20\% of CPU cycles.
Their characterization of microarchitectural fetch latency
and cache misses motivates further research in 
systematically understanding the performance of
the operational tasks.
As tasks around networking, telemetry, and security get consolidated,
sidecars offer unique opportunities for software and hardware innovations.
In this section, we outline how microarchitectural metrics
and a methodology that covers diverse policies with varying complexity
could benefit software and hardware research in service-mesh
infrastructures.

\noindent\textbf{Performance prediction and optimization in service mesh.}
Our characterization enables application operators
to reason about performance trends of diverse policies.
Inclusion of microarchitectural metrics in profilers for sidecar proxies allows us to
build automated and dynamic tools to predict and optimize 
service-mesh infrastructure for improved performance and hardware utilization.
Such tools would enable predictable service-mesh performance and improved system utilization while maintaining the desired quality of service.

\noindent\textbf{Hardware support for service-mesh infrastructures.}
Hardware vendors are designing
solutions~\cite{burres2021intel,burstein2021nvidia} 
to accelerate several cloud infrastructure components including
network and storage. 
Our characterization of the microarchitectural implications of 
sidecar proxies
with a comprehensive suite of network policies
helps architects design specialized hardware
that further offload the sidecar infrastructure efficiently.
Offloading such service-mesh components reduces 
interference, and thus enables scalable microservice applications.

\section*{Acknowledgements}
We would like to thank the anonymous reviewers for their valuable and constructive
feedback. 
This work was supported in part by ACE, one of the
seven centers in JUMP 2.0, a Semiconductor Research
Corporation (SRC) program sponsored by DARPA and by Intel RARE grant.

\bibliographystyle{IEEEtran}
\balance
\bibliography{myref}

\end{document}